


\def\di{\partial}




\def\half{{\textstyle{1 \over 2}}}

\def\L{\vec L}

\def\m{\eqno}
\def\l{\left}
\def\r{\right}

\def\Bar{\overline}

\def\R{}
\def\L{}

\def\nnn{\noindent}

\magnification=\magstep 1
\baselineskip=22pt

\centerline{\bf FLUCTUATIONS IN ISOTHERMAL SPHERES}

\vskip .1 in

\centerline{Joseph Katz\footnote\dag{E-mail: jkatz@vms.huji.ac.il}}

\centerline{The Racah Institute of Physics, 91904 Jerusalem, Israel}

\centerline{and}

\centerline{Isao Okamoto\footnote\ddag{E-mail: okamoto@yso.mtk.nao.ac.jp}}

\centerline{Division of Theoretical Astrophysics, National Astronomical Observatory, }

\centerline{Mitaka, Tokyo 181-8588, Japan}

\vskip .1 in

\noindent Sunday, December 19, 1999
\vskip .1 in

\centerline{ABSTRACT}

\vskip .1 in

Isolated isothermal spheres of N gravitationally interacting points  with equal mass are believed to be
stable when density contrasts do not exceed   $709$. That stability limit does, however, not take into
consideration  fluctuations of temperature near the onset of instability. These are important when N is finite. 

\R  Here we correlate {\it  global mean quadratic temperature fluctuations} with  onset of instability.
 We show that such
fluctuations trigger instability when the density contrast reaches a value  near
$709\cdot\exp(-3.3N^{-1/3})$. These lower values of limiting density contrasts are significantly smaller than 709 when
N is not very big and this  suggests (i) that numerical calculations with small N may not reflect correctly the onset
of core collapse in clusters with big N and (ii) that a greater number of globular clusters than is normally believed  may
already be in an advanced stage of core collapse because most of observed globular clusters whose parameters fit
quasi-isothermal configurations are close to marginal stability.\L

\nnn{\bf Keyword}: Globular clusters - Instabilities - Gravothermal catastrophe

\beginsection  1. Introduction

\R Bounded isothermal spheres of pointlike stars of equal mass interacting gravitationally are unrealistic but instructive
models to learn about the early stages of gravitational collapse in stellar clusters. Their great advantages
are  mathematical simplicity and the fact that other factors of evolution than the ``gravothermal catastrophe" like escape
of stars to infinity, equipartition of energy among stars of different masses or formation of hard binaries are
switched off by the model. Isothermal spheres have been studied in statistical mechanics (Antonov 1962, Horwitz
and Katz 1978), in thermodynamics (Lynden-Bell and Wood 1968, Katz 1978) and in numerical experiments 
(Hachisu et al. 1978, Inagaki 1980). The various studies came up with consistent results and confirmed the
following picture  of
slow evolution of isothermal spheres through different stages of global quasi-static isothermal equilibrium
with increasing entropy and density contrasts. It is nicely described  In  {\it Galactic Dynamics} by Binney and
Tremaine (1987).\L

 Consider a bounded isolated 
isothermal sphere of radius $R$ with $N$ point particles of mass $m$ attracting 
each other gravitationally. The total energy $E$ is a function of the inverse global
temperature $T$  shown in figure 1. The winding-in of pairs of thermodynamic equilibrium quantities 
like $E$ 
and $\beta=1/kT$ as shown in figure 1 is also a property of polytropic gas spheres 
(Chandrasekhar 1934), cold white dwarfs (Harrison et al. 1965), hot isentropic 
stars (Thorne 1966) and relativistic gases with $P/\rho=$constant
(Chandrasekhar 1972). Each point on the $E(\beta)$ line, like A, B, C, etc. is associated with 
an equilibrium configuration. The branch of thermodynamically stable equilibria 
indicated on figure 1 ends at point D where the density contrast is $\simeq
709$ and the energy  is also the smallest for equilibrium configurations  (Antonov 1962;
Lynden-Bell and Wood 1968).  All points of density contrasts bigger than $\simeq 709$ represent unstable 
configurations (Katz 1978). 

\R Evolution is  down the slope in figure 1 say from  point A towards point D, towards higher entropy and higher density
contrasts.  What happens to a slowly evolving isothermal
sphere when it becomes unstable?  Various interesting thought experiments have been performed and vividly 
described in Lynden-Bell and Wood  and in Binney and Tremaine. 
The system  becomes unstable when it reaches point D and a ``gravothermal 
catastrophe" develops. Notice that from B to D the entropy is a local maximum, 
while the heat capacity $C_V<0$ since the slope is positive. From D to F the 
entropy extremum is a minimum but $C_V>0$. \L 

Notice also that those thought experiments are based on the assumption that 
the system reaches a new local entropy minimum in a {\it finite} time. In fact the 
system spends a limited time at each point of the linear series in 
(quasi-)equilibrium. It is thus assumed that during the evolution, 
\R fluctuations, and in particular  mean global temperature fluctuations (which we shall precisely define
in a moment)\L are small, i.e. that the probability of a big  fluctuation is negligibly small. This
assumption is certainly not correct near  point D where fluctuations  of  global temperature become
infinite (Horwitz and Katz 1977 referred to below as HK77). One of the main objects of the  present work is to try
to assess how close to point D  the evolution can proceed  before the system becomes unstable as a result of big
fluctuations. 
\R If
the system spends an unlimited  time in a stable equilibrium  with energy less negative than $E_D$, it
will eventually evolve to a non isothermal state because big fluctuations of  temperature 
have a chance to develop and bring the system to a state of entropy which would  increased indefinitely (see below) . \L

\R  Now a few words about temperature fluctuations in isothermal spheres which are at the center of our
considerations. The global temperature T is a global equilibrium parameter but it has also a local meaning 
(see Lynden-Bell and Wood again); T is related to the mean square velocity $\Bar {v^2}$ by the same relation as
in an ideal gas
${\textstyle{3\over 2}}kT=\half m\Bar {v^2}$. The mean square velocity is that of the stars which happen to
be near a point at a given moment. At later times stars left the vicinity and other stars came
around. A local temperature can also be defined  if the cluster is not in global thermal equilibrium but
such a local temperature would differ from point to point and from time to time. A global equilibium
temperature does then not exist. The proper method to tackle the problem is then in kinetic theory not statistical
mechanics. \L
 
\R However, if an out of equilibrium entropy can be defined in statistical physics for a system with a given
energy $E$ [see HK 1977], say $w$, than a global out of equilibrium temperature
$\tilde{T}$ can be defined as well by
$\di w/\di E=1/k\tilde{T}=\tilde{\beta}$. That $\tilde{T}$ will clearly be equal to $T$ at the point where $w$ is equal to its
maximum $S$ and fluctuations of $T$ can be calculated like in classical statistical physics (see Landau and
Lifshitz 1985).\L Thus we may state without proof (but we shall  give the prove  below) that the mean
quadratic fluctuations of the global temperature $\Bar {(\Delta T)^2}/T^2$ (from now on temperature will
always mean global temperature)  is, like in classical thermodynamics,   of
order 
$1/N$. Thus, if $N$ is big, $\Bar {(\Delta T)^2}/T^2$ is small and  fluctuations may be expected
to remain very small along the linear series up to or rather very near  the stability limit at point D. In
gravitational thermodynamics $N$ is much  smaller than in atomic physics; $N\sim 10^{10}$ in
galactic nuclei, 
$10^{5\sim 6}$ in globular clusters and $10^{3\sim 4}$ in present day numerical 
simulations (Meylan and Heggie 1997). Fluctuations in these different cases  are wildly different and some have measurable
effects as we shall see.

In this connection, it is worthwhile  recalling Monaghan's (1978) application 
of the theory of {\it hydrodynamic} fluctuations to a self-gravitating gas. 
He showed  that density fluctuations become large before the point of ordinary 
stability is reached. Statistical mechanics confirms Monaghan's finding. 

 In the present work  we apply our own version  of  
fluctuation theory in $N$ body systems in statistical physics with gravitational
forces (Parentani et al. 1995). The theory, first sketched in
Okamoto  et al. (1995), has been applied to evolutionary sequences of quasi-static  equilibrium
configurations of self-gravitating radiation in the presence of  a black hole. Such systems exhibit a 
first order phase
transition and fluctuation theory  plays a useful role in explaining when superheated black holes or
superheated  radiation become unstable. 

Fluctuation theory plays a very different role in an evolutionary
sequence of quasi-static isothermal spheres, for instance.  These are in a state of {\it
local} 
entropy maximum and the maximum is unique.\R The global entropy maximum  is in fact infinite as shown by Antonov. The
non-equilibrium entropy has thus a local maximum and a local minimum. It is qualitatively like the dashed curve in figure 2.
Large enough fluctuations in equilibrium configurations near marginal stability at point D may bring the system along the
dashed line at a point where the entropy can grow towards the global maximum, at infinity. 
 Our aim  is  to find a limit where {\it global temperature fluctuations} become  large enough to put the system  out of
equilibrium. Such a limit exists if the mean square amplitude of temperature fluctuations grows
exponentially. This is  the case in general and as we shall see it is certainly so in isothermal spheres.\L

Our theory of fluctuations  applies not only to isothermal 
spheres but to any thermodynamic equilibrium with long range forces which 
has a {\it unique} but local minimum of entropy and satisfy  some additional conditions which we describe in
detail below. 

In section 2 we review the theory of fluctuations that has been presented in
 Parentani et al. (1995) 
and in section 3 we apply the theory to isothermal spheres. As we shall see, 
the stability limit induced by fluctuations depends  exponentially  on a power of $N$ and the density contrast at which
instability develops may be significantly 
lower than 709 for ``small" $N$. The significance of these calculations is analyzed in section 4.

\beginsection 2. Fluctuation theory with long range forces

\noindent (i) {\it Steepest descent expression for the entropy of a finite isolated 
system}

Consider a system of gravitationally interacting particles in a finite volume 
$V$, with total mass $M$ and fixed energy $E$. The total Gibbs entropy is 
$S=k\ln\Omega$, where $\Omega$ is the volume of available phase space; it 
can be evaluated by a steepest descent method, the saddle point value giving 
the mean field entropy (HK77).  
Analysis of the quadratic fluctuations yields thermodynamical stability conditions 
 and the next order term in calculating the entropy 
 beyond the mean field approximation (see for instance Horwitz 1971). For a 
recent review on the subject of statistical mechanics of gravitating systems, 
see Padmanabhan (1990). In HK77's scheme $\Omega$ is given by a functional 
integral but one needs only consider continuous functions (Ginibre 1971) 
and, under general conditions (Courant and Hilbert 1953) which are usually met 
in physical problems, the continuous functions can be expanded in an 
eigenfunction series which converges absolutely and uniformly in the whole 
domain of existence. Under such circumstances, the functional integral can be  
replaced by discrete integrations in terms of an infinite but denumerable set 
of variables say $x^{i}$. $\Omega$ assumes thus a form like this 
$$
\Omega(V,M,E)={1\over A}\int_{L\to\infty}e^{w(V,M,E;x^{i})}dx^1dx^2...dx^{L}.
\m (2.1)
$$
In the steepest descent evaluation, $w$ is extremized, 
$$
{\di w\over \di x^i}=0.
 \m (2.2)
$$
 $w$ is then expanded in powers of $(x^i-X^i)=\Delta x^i$ where $X^i(V,M,E)$ is a
solution of (2.2). We shall assume that the $x^i$'s are chosen in such a way
that at the point of extremum indicated by a sub-index $e$, the matrix of second 
derivatives of $w$ is diagonal; thus 
$$ 
\left({\di^2 w\over\di x^i\di x^j}\r)_e =-\delta_{ij}\lambda_j.
\m (2.3)
$$
The minus sign is for convenience{\footnote\ddag{By assuming that variables may be chosen such that (2.3) 
holds, we bypass of course the main technical problem of stability theory 
which is {\it to find} the $\lambda_i$'s...}. We can always denumerate the $\lambda_i$'s in 
such a way that $\lambda_1\leq\lambda_2\leq\lambda_3...$~. 
.
Expansion of $w$ 
in powers of $\Delta x^i$ may thus be written as follows (if $A$ is properly chosen)
$$
w={1\over k}S(V,M,E) -{1\over 2} \mathop{\sum}^{\infty}_{i=1}\lambda_i(\Delta x^i)^2 +O_3. 
\m (2.4)
$$ 
The $\lambda_i$'s are known as Poincar\'e's coefficients of stability following  Poincar\'e's 
(1885) seminal analysis of stability. In writing (2.4) we have nowhere assumed that we deal with 
pure or not pure gravitational interactions. (2.4) is therefore
quite general and applies to non-extensive systems. 

Equation (2.4)  cries for
qualifying restrictions: 

(1) The summation like the integral 
is in general not convergent (see HK77 for details). What makes them  converge is 
a short distance cutoff in the gravitational interaction. The effect of the cutoff 
is to modify the higher values in the spectrum $\lambda_i$ which 
play no role in stability and fluctuation considerations  (see below). The effect of  short distance
cutoffs on stability was analyzed in Aronson and Hansen (1972).

(2) $w$ must admit second order derivatives with 
respect to  $x^i$. We shall assume that $w$ is twice differentiable also with 
respect to $V$, $M$ and $E$ in the vicinity of the maximum otherwise thermodynamic coefficients would not exist. In what
follows we shall concentrate on the
$E$  dependence of equilibrium configurations and take $V$ and $M$ fixed. Therefore 
we write $S(E)$, $\beta (E)$ etc... 

\R(3) We shall now regard  the smallest $\lambda_i$'s to be different from each other. This assumption is more than we shall
need and is made for convenience; it simplifies explanations. In the paper by Horwitz and Katz of (1978) it was shown that in
stable isothermal spheres $\lambda_1>0$ and that at density contrasts near but greater than 709,  $\lambda_1<0$ but
$\lambda_2>0$. Another case in which a few of the smallest
$\lambda_i$ have been shown to be  different is in models for stars of cold catalyzed matter; this includes cold
neutron stars (see in Harrison et al. 1965). At each turning point of the linear series $-\mu (N)$ where $\mu$ is
the chemical potential\footnote\S{ The curve does not appear in Harrison et
al.; it is represented schematically in Katz (1981)  and is based on Harrison et al.'s tables.} $\mu=dE/dN$  there is a
change of stability at each of the three turning points calculated and each corresponds to {\it one} and only one square
eigenfrequency ($\omega^2$) of dynamical perturbations  changing sign.   Degenerate spectra of $\lambda_i$'s are in fact
rare in astrophysical applications though not unknown. Degeneracy corresponds to bifurcations of  equilibria; see for
instance the many bifurcations in liquid  ellipsoids linear series (Chandrasekhar 1969). They usually
correspond to an excess of symmetry which can be lifted with small arbitrary perturbations 
of the  potential energy [see Thompson (1979), for Maclaurin ellipsoids in particular see Katz (1979)].
There are a few exceptions of physical interest in which bifurcations cannot be removed (see Arnold
1986). When all we are interested in is the passage of stability to instability   it is enough for  $\lambda_1$ to be
different from $\lambda_2$. Some interesting results about the onset of instability  may also be obtained without
assuming that the spectrum of $\lambda_i$'s is non degenerate (Sorkin 1981). \L

\noindent (ii) {\it The standard limit of stability}

\R With (2.4), we now define a ``temperature function" $\tilde{T}(E;x^i)$ 
or rather the more convenient inverse temperature or $\tilde{\beta}=1/k\tilde{T}$  which has the same units 
as $E^{-1}$:
$$
\tilde{\beta}={\di w\over \di E} 
  ={1\over k}{dS\over dE}+\mathop{\sum}_{i}\lambda_i{dX^i\over dE}\Delta x^i +O_2 
\equiv \beta +\mathop{\sum}_{i}\lambda_i{dX^i\over dE}\Delta x^i +O_2.
\m (2.5) 
$$   
The extremal value of $\tilde{\beta}$ is $\beta=1/kT$, $T$ is {\it the} 
global temperature of equilibrium. For small values of $\Delta x^i$'s \L
$$
 \Delta\tilde{\beta}=\tilde{\beta}-\beta \simeq \mathop{\sum}_{i}\lambda_i{dX^i\over dE}\Delta x^i. 
\m (2.6)
$$
These formal fluctuations are related to fluctuations of
temperature in a  small part of the system in the usual sense when the rest of the system can 
be treated as a reservoir [see Parentani et al. (1995) for a detailed discussion 
of this point]. That, however, is not normally the case in gravitating systems.

Since $w$ is twice differentiable, derivatives of $\tilde{\beta}$ with respect 
to $x^i$ and $E$ exist; the derivative of $\tilde{\beta}$ defined in  (2.5) 
with respect to a certain $x^i$, keeping $E$ constant and all other $x^i=X^i$ 
is readily derived from (2.6) 
$$
{\di{\tilde{\beta}}\over \di{x^i}}=\lambda_i{d{X^i}\over dE}=\l({\di{\tilde{\beta}}\over \di{x^i}}\r)_e~~~[{\rm no~summation~on}~ i]
\m (2.7)
$$
and the derivative of $\tilde{\beta}$ with respect to $E$, keeping {\it all} $x^i=X^i$ is obtained from (2.5) 
$$
{\di\tilde{\beta}\over \di E}={d\beta\over dE}-\mathop{\sum}_{i} \lambda_i\l({dX^i\over dE}\r)^2 
\m (2.8)
$$
Now replace $dX^i/dE$ in (2.8) in terms of $({\di\tilde{\beta}/ \di x^i})_e$ 
as given by (2.7); then ${\di\tilde{\beta}/ \di E}$ can be written 
$$
{\di\tilde{\beta} \over \di E}={d\beta\over dE} -\mathop{\sum}_{i} {1\over\lambda_i} 
      \l({\di \tilde{\beta}\over \di x^i}\r)_e^2 . 
\m (2.9)
$$
We now turn our attention to the {\it stability limit}. The entropy is a
maximum and the $X^i(E)$'s represent a stable thermodynamic mean field configuration  
when all the Poincar\'e  coefficients are positive, i.e. when $\lambda_1>0$. Consider thus a 
linear series $\beta (E)$ of stable equilibrium configurations. Any change of stability or change in the
degree of instability  is characterized by a change of sign in one of the $\lambda_i$'s. \R Thus when a
series of {\it stable} equilibria becomes unstable, $\lambda_1$  becomes small and   changes sign.
 Near instability, the sum in (2.9) is   dominated by the $1/\lambda_1$ term {\it assuming} $\l({\di\tilde{\beta} / \di x^1}\r)_e\neq 0$. If  $\l({\di\tilde{\beta} / \di x^1}\r)_e = 0$ we can get rid of the coincidence of
zeros by adding a small perturbation $\delta w(x^i)$ to $w$, for instance $\epsilon x^1E$ with $\epsilon$ small enough. Then
$\l({\di\tilde{\beta} / \di x^1}\r)_e = \epsilon$.  
This is the normal way bifurcations are removed in ordinary stability analysis (Hunt 1977)\footnote\ddag{Notice however
that we do not know what the $x^i$'s are in general to begin with (see footnote p.7). For details on this point see Katz
(1980).}.  Small perturbations will 
generally get rid of such degeneracies (but not always). We  assume that 
$\l({\di\tilde{\beta} /
\di x^1}\r)_e\neq 0$ or, if necessary, is non zero with a small perturbation of
$w$ . If the sum in (2.9) is indeed dominated by the first term which diverges when
 $\lambda_1=0$, and if both $\di\tilde{\beta}/\di  E$ and $(\di\tilde{\beta}/\di x^i)_e$ are finite,
then for $\lambda_1\to 0$,  $d\beta/dE$ must  diverge  and it follows from (2.9) that\L
$$
{d\beta\over dE}\simeq {1\over\lambda_1} \l({\di \tilde{\beta}\over\di x^1}\r)^2_e 
\m (2.10a)
$$
or that
$$
\lambda_1\simeq
{dE\over d\beta}\l({\di \tilde{\beta}\over\di x^1}\r)^2_ e.
\m (2.10b)
$$
This is Poincar\'e's turning point property of $E(\beta)$: the energy is maximum or 
minimum when a change of stability takes place. With (2.10) we can calculate the heat capacity $C_V$ in stable configurations
{\it near instability}:

$$
C_V ={dE\over dT}\simeq -k\lambda_1\beta^2\l({\di\tilde{\beta}\over\di x^1}\r)_e^{-2}<0.
 \m (2.11)
$$
Thus $C_V<0$ for a stable equilibrium $(\lambda_1>0)$ and $C_V>0$ for an unstable 
one. This property is true near a turning point. It is interesting to
remind the reader that 
$C_V<0$ in stable systems is a property  of 
{\it finite} non-extensive microcanonical ensembles in thermal equilibrium and not  of gravitating systems only. Notice that
a  microcanonical ensemble is here stable where a  canonical ensembles is unstable  ($C_V<0$). 

\noindent (iii) {\it Mean quadratic fluctuations of temperature and their probability} 

Consider now a stable equilibrium configuration near a point of instability
where 
$\lambda_1(>0)\to 0$ and $\lambda_i>0$ for $i\geq 2$. All terms of the quadratic sum 
in (2.4) except the one with $\lambda_1$ may be integrated out in 
(2.1); they are by definition "strongly¾ stable if they have
 strongly negative 
exponentials which is generally the case when $N>>1$. Thus, near instability, we are interested in the terms of $w$
that 
are not (strongly) stable, i.e.\ in 
$$ 
 w\simeq{S(E)\over k}-\half \lambda_1(\Delta x^1)^2
 \m (2.12)
$$
with a slightly  renormalized factor $A$ in (2.1). We may 
replace $\lambda_1$ in (2.12) by its expression given in (2.10b).
Near instability, {\it with all other $x^i$'s integrated out}, the (new) 
temperature function $\tilde{\beta}$ calculated from $w(E;x^1)$ reduces to 
$$
\tilde{\beta}=\beta+\Delta \tilde{\beta}\simeq\beta+\l({\di\tilde{\beta}\over\di x^1}\r)_e \Delta x^1.
\m (2.13)
$$
If we now replace $\Delta x^1$ in (2.12) in terms of $\Delta\tilde{\beta}$ given by (2.13) we obtain
$$
w\simeq {S(E)\over k}-\half {dE\over d\beta}(\Delta \tilde{\beta})^2. 
\m (2.14)
$$
Equation 
(2.14) is of the same form as equation (110.3) of Landau and Lifshitz's (1985) chapter on (non-quantum) 
fluctuation theory. We can now  use the standard arguments of fluctuation theory and say that the probability 
$dW$ for a fluctuation of $\tilde{\beta}$ in the range, $\beta+\Delta\tilde{\beta}$ and 
$\beta+\Delta\tilde{\beta}+d\tilde{\beta}$, is proportional to $\exp(w-S)$; thus, the properly 
normalized  $dW$ is given by 
$$
dW=\sqrt{{1\over2\pi}{dE\over d\beta}}\exp\l[-\half{dE\over d\beta}(\Delta\tilde{\beta})^2 \r] d\tilde{\beta} 
\m (2.15)
$$
From (2.15) it follows that the mean quadratic fluctuations of the global temperature,
as a function of the equilibrium parameters, is given by 
$$
\Bar {(\Delta\tilde{\beta})^2}={d\beta\over dE} =-{k\beta^2\over C_V}. 
\m (2.16)
$$ 
The analogy with classical fluctuation theory is flagrant except for the sign of $C_V$. 
Notice that our analysis does 
not use small subsystems treating the rest of the system as a heat bath. Small subsystems have no well defined
energies independently of the 
 big system when there are long range forces. The intriguing difference of signs of $C_V$ and comparison with 
familiar results is discussed in detail in Parentani et al. (1995).
Before getting any further it is good to remember that (2.15) and 
(2.16) are valid only  close to the point of marginal stability, point D, where 
$dE/d\beta=0$, $C_V=0$ and $\Bar{\Delta \tilde{\beta}^2}=\infty$. The derivative of $E(\beta)$ near point D may  be
approximated by the lowest order in a Taylor expansion:
$$ 
{dE\over d\beta}\simeq \l({d^2E\over d\beta^2}\r)_D(\beta-\beta_D).
\m (2.17)
$$
and with (2.17) the exponential factor in $dW$ can be written 
$$
\exp\left[-\half\left(\beta^3{d^2E\over
d\beta^2}\right)_D\l({\beta-\beta_D\over\beta_D}\r)\l({\tilde{\beta}-\beta\over\beta_D}\r)^2
\r].
\m (2.18)
$$
There appears here two dimensionless quantities which deserve a special symbol, say
$$ 
\Delta U={\beta-\beta_D\over\beta_D} ~~{\rm and} ~~\Delta \tilde{V}={\tilde{\beta}-\beta\over\beta_D} 
\m(2.19)
$$
 $\Delta U$ depends on equilibrium parameters only, 
while $\Delta \tilde{V}$ is a fluctuation in units of $\beta_D$. \R The term in 
(2.18) that contains second order derivatives of $E$ is non-dimensional. The virial
theorem for particles in a container tells us that the total kinetic energy and the total
potential energy are of the same order of magnitude. Thus the total energy E is itself  of the same order of magnitude
than the total kinetic energy. On the other hand $kT=1/\beta$ is of the order of magnitude of the kinetic
 energy of one particle.  Therefore the first factor in 2.18, which ``goes like $\beta E$" must be of order N \L
$$
\l(\beta^3{d^2E\over d\beta^2}\right)_ D=\Gamma_D N .
\m (2.20)
$$
where the number $\Gamma_{\rm D}$ must be of order $1<<N$. In terms of $\Delta U$, 
$\Delta\tilde{V}$ and $\Gamma_ D$, the probability $dW$ defined in (2.15) 
can be written 
$$
dW=\sqrt{{\Gamma_D\over2\pi}N\Delta U}\exp{\l[-\half\Gamma_D N\Delta U (\Delta\tilde{V})^2\r]} d\tilde{V} .
\m (2.21)
$$
The expression for the mean quadratic fluctuations (2.16) can be rewritten in terms of $U$'s and $\tilde{V}$'s in the
following form
$$
 \Bar{(\Delta \tilde{V})^2}=(\Gamma_D N \Delta U)^{-1}. 
\m (2.22)
$$
\R 
The rate at which a small fluctuation of temperature 
disappears is, by virtue of the fluctuation-dissipation theorem,
 the inverse of the time it takes to return to equilibrium, $1/ t_{rel}$. This relaxation time is proportional to
some
 power of N (see for instance Binney and Tremaine about stellar systems). But $dW$ is a power of $e^{-N\Delta U}$
and \L 
 if $\Delta U$ is not very small (say $\Delta U\sim 0.1$), the exponent becomes extremely 
small for large $N$( say $N>100$). The probability per unit time dW/dt will thus 
also  be dominated by the negative exponential of $dW$.

\noindent (v) {\it The destabilizing effect of fluctuations near a local maximum of 
entropy}  

Let us now determine the role of global temperature fluctuations near marginal stability. Consider a sequence of
quasi-equilibrium 
configurations evolving with a relative rate of change in volume
$V(dt/dV)$ of a few $t_{rel}$. The quasi-static linear series has a local entropy 
maximum and a local entropy minimum (see figure 2). The evolutionary sequence   approaches  the turning 
point D through  a succession of quasi-equilibria. \R
However, as point D is approached, the mean quadratic fluctuations tend to infinity as can been seen from (2.16) since
$C_V
\rightarrow 0$ and the system is unlikely to remain in thermal equilibrium. So how close to point D
will the system survive in quasi-equilibrium?
\R Let us try to define such a point, call it C, and find a posteriori if it has any physical sense. 

Let $T_C$ be the temperature at point C and $\Bar{(\Delta \tilde{T})^2}_C$ its mean quadratic fluctuation. We define point C
as one at which (see figure 2) the mean quadratic fluctuation is just equal to $(T_E-T_C)^2$. For smaller {\it mean}
fluctuations big {\it real} fluctuations have only a negligible probability to put the system in a state of  entropy
at the left hand side of the minimum of entropy in figure 2 (in one relaxation period). If mean  quadratic fluctuations are 
bigger than
$(T_E-T_C)^2$  the probability for a real big fluctuation in a relaxation period increases also.  Big fluctuations may  bring
the system in a state in which the entropy can increase indefinitely and if it can it will so that thermal equilibrium will
 be lost. Point C defines a limit between these different behaviors; at C
$$
\sqrt{\Bar{(\Delta \tilde{T})^2}_C}= T_E-T_C \simeq 2(T_D-T_C)
\m (2.23)
$$ 
The last quasi-equality comes from a quadratic approximation near the horizontal tangent. In terms of $U$'s and $\tilde{V}$'s defined in (2.19), (2.23) can be written    
$$
[\Bar{(\Delta \tilde{V})^2}]_C =[2(\Delta U)_C]^2 . 
\m (2.24)
$$
and if (2.24) holds then, following  (2.22), 
$$
(\Delta U)_C =(4\Gamma_D N)^{-1/3}
\m(2.25)
$$
This equality defines $\Delta U_C$ and thus $\beta_C$ and also $T_C$. At point C
the exponent in (2.18) is just equal to
$$
-{\textstyle{{1 \over 8}}} \l({\Delta\tilde{V}_C\over \Delta U_C}\r)^2
=
-{\textstyle{{1 \over 8}}} \l({\tilde{\beta}-\beta_C\over\beta_D-\beta_C}\r)^2
=
-{\textstyle{{1 \over 8}}} \l({\tilde{T}-T_C\over T_D-T_C}\r)^2.
\m (2.26)
$$
(2.25) gives the following value for $\beta_C$
$$
\beta_{\rm C}=\beta_D[1+(\Delta U)_{\rm C}]=\beta_D [1+(4\Gamma_D N)^{-1/3}].
\m (2.27a)
$$
The corresponding value of the energy $E_C$ can be estimated by expanding 
$E(\beta)$ near point D in powers of $\beta_C-\beta_D$, remembering that $(dE/d\beta)
_D=0$. Taking account of (2.20), we have, to  second
order in $\beta_C-\beta_D$, using also (2.27a):\L
$$
E_C\simeq E_{\rm D}\l[ 1+{(\Gamma_D/2)^{1/3}\over4(\beta E/N)_{\rm D}} N^{-2/3} \r]
\m(2.27b) 
$$
$(\beta_C, E_C)$ are the coordinates of point C.

\beginsection{3. Applications to Isothermal spheres}   

Isothermal spheres have been studied in great detail by many people. We refer 
to Binney and Tremaine for a modern presentation of the theory of isothermal 
spheres. With a Maxwellian distribution of energy per particle, the density 
$\rho(r)$ depends exponentially on the gravitational potential $\rho\sim 
e^{-\beta\Phi}$. This explains why Newton's equation can be written in the 
form [see equation (371) in Chandrasekhar (1934) or (4.15b) in Binney and 
Tremaine] 
$$
{d\over dr}\l(r^2{d\ln\rho\over dr}\right)=-4\pi Gm\beta r^2\rho.
\m(3.1)
$$
The equation can be replaced by a set of two first-order  differential
equations 
(Emden 1907). Here we use the $(u,v)$ variables introduced by Chandrasekhar; first define the mass within a radius $r$ as
$\mu(r)$ 
$$ 
\mu =\int^r_0 4\pi\rho r^2 dr. 
\m (3.2)
$$
Then introduce the variables 
$$
 u={4\pi\rho r^3\over\mu}, \quad 
  v= 4\pi Gm\beta \rho r^2 \quad {\rm and} \quad
  \zeta=\ln r. 
    \m (3.3)
$$

Equation (3.1) is equivalent to the following pair of equations for 
$u$ and $v$ (Chandrasekhar's equations (404) and (405))
$$
{du\over d\zeta}=u(3-u)-v, 
\m(3.4)
$$
$$ 
{dv\over d\zeta}= 2v-{v^2\over u} \quad . 
\m(3.5)
$$
Tables of solutions of these equations have been published by Emden (1907) 
and Chandrasekhar and Wares (1949). The total energy $E$ of isothermal spheres 
is calculable in terms of $u_B$ and $v_B$, the boundary values at $r=R$ of $u$ 
and $v$. 
The non-dimensional total energy and inverse temperature in figure 1 are the following functions of $u_B$ and $v_B$; 
$$ 
E^* = \l({R\over GNm^2}\r){E\over N}
   ={u_B\over v_B}(u_B-{3\over 2}) 
\m (3.6) 
$$
and from (3.3) alone, since $\mu(R)=\mu_B=M=Nm$,
$$
b ={GNm^2\over R}\beta ={v_B\over u_B}.  
\m (3.7)
$$ 
Reciprocally, in terms of $E^*$ and $b$  
$$ 
u_B=bE^*+{3\over2}, \quad  v_B=b\l(bE^*+{3\over 2}\r). 
\m (3.8)
$$

From (3.8), (3.7) and from (3.4), (3.5)  at $r=R$ we obtain a differential equation for the linear series $E^*(b)$
represented in figure 1: 
$$ 
{dE^*\over db}=-{1\over b^2}\l( 2bE^*+ b-\half  +{b-2\over bE^*+ \half} \r).
 \m (3.9)
$$
Equation (3.9) can be integrated directly; however the parametric 
form (3.4)-(3.5) deduced from (3.6)-(3.7) is more convenient for numerical integration; with $\zeta_B=\ln R$ we have
$$
{dE^*\over d\zeta_B}
=
-{1\over b}
\l[2(bE^*+\half)^2 + (b-{\textstyle{3 \over 2}})(bE^*+\half) + b -2\r],
\m (3.10)
$$
$$
{db\over d\zeta_B} =b\l(bE^*+ \half\r).
\m (3.11)
$$ 
The density contrast $\cal R$ is obtained by integrating twice equation 
(3.1). In terms of $u$, $v$ and $\zeta$, a first integration gives  
$$
{d\ln\rho\over d\zeta}=- {v\over u} . 
\m (3.12) 
$$
With (3.7), (3.11) and (3.12) it then follows that 
$$
{\cal R}={\rho (0)\over\rho(R)}=\exp\int^{b}_{0}{dx\over xE^*(x)+\half}
\m (3.13) 
$$
Figure 1 has been obtained by integrating  (3.10) and (3.11) 
setting $\zeta=0$ at a point with $b_0=1.549$ and $E_0\simeq 0.327$ 
calculated from Emden's tables. 

Let us now evaluate the quantities related to the stability limit triggered by 
 fluctuations. Point D, has coordinates $b_D$, $E^*_D$ and a density contrast ${\cal R}_D$ all equal to
$$
b_D= 2.03, \quad E^*_D= -0.335, \quad {\cal R}_D=709. 
\m (3.14)
$$ 
The number $\Gamma_D$ associated with point D, and defined in (2.20) can be derived from (3.9); it is non-dimensional
and can be written using (3.6) and (3.7) like this 
$$
\Gamma_D=\l(b^3{d^2 E^*\over db^2}\r)_D =-b_D\l(2E^*_D+1 +{2E^*_D+\half\over(E^*_Db_D+\half)^2}\right)\simeq 9.95\approx
10
\m (3.15)
$$
Thus, following (2.27), the coordinates of the stability limit due to fluctuations are
$$ 
b_C \simeq b_D (1+\Delta b_C/b_D) = 2.03(1+0.29 N^{-1/3})
\m (3.16a)
$$
$$
E^*_C\simeq E^*_D(1+\Delta E^*_C/E^*_D) =-0.335(1-0.63N^{-2/3})
\m (3.16b)
$$ 
at which point the density contrast, according to (3.13), is 
$$
{\cal R}_C= {\cal R}_D \exp\int^{b_C}_{b_D}{dx\over xE^*(x)+\half} 
\simeq {\cal R}_D \exp\l({\Delta b_C\over b_CE^*_D+\half}\r) 
\m  (3.17)
$$
Thus, taking account of (3.16)  we obtain 
$$
 {\cal R}_C\simeq  709\cdot\exp{(-3.30 N^{-1/3})} 
\m(3.18) 
$$

\beginsection {4.  Remarks on these results and some observational implications}

The Table  gives the stability limit induced by fluctuations for values of $N$ in the range $10\leq N\leq\infty$. 
In addition to the coordinates $(b_C,E^*_C)$ and the corresponding values of 
density contrasts ${\cal R}_C$, the table also provides 
relative changes in temperature $\Delta b_C/b_D$ and energy $\Delta E^*_C/E^*_D$ which give  
indications about the validity of the linear approximation near point D. 
Notice that relative corrections of $b_C$ and $E^*_C$ do not exceed $15$\% for $N$ as small 
as $10$.  The relative reduction of the density contrasts $\Delta{\cal R}_C/{\cal R}_D$, on the other hand changes 
significantly. 

The numbers have a simple interpretation: in isolated sphere in slow quasi-static evolution towards higher 
and higher density contrast as described in the Introduction the
gravitational catastrophe or core collapse  appears  at lower density contrasts than $709$. The change from 
${\cal R}_D$ to ${\cal R}_C$ is very small for $N= 10^{6\sim 5}$ but quite significant 
for $N\simeq 10^3$ and becomes  drastic at lower values of $N$. 

\R A gravothermal catastrophy, as is well known, is not like an avalanche; the central parts of the stellar
system gets hotter while the outer parts are left behind [see Lynden-Bell (1999) for a recent revue on gravothermal
catastrophy]. The instability  induces  a change in the evolution which accelerates progressively. Our  point 
here is to note that the change does not happen at point D but rather earlier, at point C. \L 

How sharply is point C defined?
To appreciate the sharpness of this new  stability limit, 
consider the probability distribution $dW$ at an inverse (non-dimensional) 
temperature $b>b_C$. Consider also a fluctuation $(\tilde{b}-b)$ (refere to figure 1 again) as 
big as $2(b-b_D)$, big enough to put the system in a state of ever 
growing entropy.  The probability distribution for such a fluctuation defined by 
(2.21) can here be written in terms of $b$'s rather than $U$'s; we obtain 
for $dW$: 
$$ 
dW=.465N^{1/3}r^{1/2}\exp{(-\half r^{3})} d\tilde{V} 
\m (4.1)
$$ 
where 
$$
r={b-b_D\over b_C-b_D} 
\m(4.2)
$$
Thus for a given $r$
$$ 
b=b_C+(r-1)(b_C-b_D) .  
\m (4.3)
$$
For $N=10^{6}$ (or $N=10^{2}$), at point $b\simeq 1.01$ ($1.12$) higher 
than $b_C$  with a factor $r=3$, the exponential factor $\exp {(-\half
r^{3})}\simeq 10^{-6}$ already. At $b \simeq 1.15$  ($1.18$) higher than $b_C$ with a factor $r=4$ , $\exp {(-\half
r^{3})}\simeq 10^{-14}$. Thus, the probability (per unit time)  
for  fluctuations to be big enough to render the system unstable changes cery steeply
for small departures from $b_C$. {\it In this respect, point C  is   rather  sharply defined }. 

The gravothermal catastrophe must thus take place at lower density contrasts because of the fluctuations and this should 
show up in
$N$-body  calculations with isothermal spheres evolving through quasi-static 
configurations.  Possible implications for
real 
globular clusters are as follows.  Galactic globular clusters are of course not 
closed isothermal spheres.  They are correctly modeled by Michie (1963)-King 
(1966) models which have truncated Gaussian distributions and are not in 
thermodynamic equilibrium. However to the extent that our results for
isothermal  spheres have any bearing on the behaviour of observed globular clusters, we speculate that a greater 
number of globular clusters appearing in the tables of Trager et al.
(1993) than indicated there may actually be in an advanced stage of core collapse because most observed globular clusters
whose parameter allows fitting them to energy truncated gaussian distributions are close to marginal stability, that is to
point D (Katz 1980).

\beginsection {\bf Acknowledgements}

 J.K. is grateful for the kind hospitality of the Division of Earth Rotation and
Mizusawa Astrogeodynamics Observatory in Mizusawa, where much of this work was done 
in a peaceful and inspiring setting. He also thanks very much Donald Lynden-Bell for 
clarifying discussions when we were at the Department of Theoretical Physics of Charles 
University in Prague.  Gerald Horwitz of the Racah Institute helped  to understand  the 
valuable clarifications demanded by the second referee; many thanks to both of them.

\centerline {\bf Bibliography}

\noindent Antonov V.A. 1962,  Vestnik Lenigrad Univ. {\bf 7}, 135 [translated in english in   {\it Dynamics of Star Clusters}
 1985, eds. Goodman J. and Hut P. , IAU Symposium No 113  (Reidel, Dordrecht)
 
\noindent Arnold V.I. 1986, {\it Catastrophe Theory}, 2nd ed, (Springer Verlag, Berlin, Tokyo)

\nnn Aronson E.B. and  Hansen C.J. 1972,  Astrophys.J. {\bf 177}, 145 

\nnn Binney J. and Tremaine S. 1987,  {\it Galactic Dynamics} (Princeton University Press, New Jersey) 

\nnn Chandrasekhar S. 1934, {\it An Introduction to the Study of Stellar Structure} (Dover 1961, New York)  

\nnn Chandrasekhar S. 1969, {\it Ellipsoidal Figures in Equilibrium} (Yale Univ. Press, New Haven)

\nnn Chandrasekhar S. 1972, in {\it General Relativity - papers in honor of J.L. Synge} Ed. $~$ O'Raifearthaigh OUP, Oxford)

\nnn Chandrasekhar S. and Wares G.W. 1949,  Astrophys.J. {\bf 109} 551

\nnn Courant R. and Hilbert D. 1953, {\it Methods in Mathematical Physics} Vol. 1, (Interscience, New York)

\nnn Emden R. 1907,  {\it Gask\"ugeln} (Leipzig) 

\nnn Ginibre J. 1971 in {\it Statistical Mechanics and Quantum Field Theory} eds.
DeWitt C. and Stora R. (Gordon and Breach, New York)

\nnn Harrison B.K., Thorne K.S., Wakano M. and Wheeler J.A. 1965, {\it Gravitation Theory 
and Gravitational Collapse} (University Press, Chicago) 

\nnn Hachisu I., Nakada Y., Nomoto K. and Sugimoto D.   1978, Prog. Theor. Phys. {\bf 60} 393

\nnn Inagaki S. 1980, Publ. Astr. Soc. Japan {\bf 32} 213 

\nnn Horwitz G. 1971, J. Math. Phys. {\bf 14} 658

\nnn Horwitz G. and Katz J. 1977, Astrophys.J. {\bf 211} 226 (referred to as HK77)

\nnn Horwitz G. and Katz J. 1978, Astrophys.J. {\bf 222} 941

\nnn Hunt G.W. 1977 Proc. R. Soc. London {\bf A 357} 193

\nnn Katz J. 1978, MNRAS {\bf 183} 765

\nnn Katz J. 1979, MNRAS {\bf 189} 817   

\nnn Katz J. 1980, MNRAS {\bf 190} 497   

\nnn Katz J. 1981, {\it Linear Series and Catastrophies in Astronomy} in Proceedings of the 5th G\"otingen-Jerusalem
Symposium (Vandenhock, Ruprecht)

\nnn King I.R. 1966, Astrophys.J. {\bf 71} 64

\nnn Landau L.D. and Lifshitz E.M. 1985,  {\it Statistical Physics} 3rd ed. (Pergamon Press, Oxford)  

\nnn Lynden-Bell D. 1999 Physica {\bf A263} 293

\nnn Lynden-Bell D. and Wood R. 1968, MNRAS {\bf 138} 495

\nnn Meylan G. and Heggie D.C. 1997, Astron. Astrophys. Rev. {\bf 8}  1

\nnn Michie P.W. 1963, MNRAS {\bf 138} 495

\nnn Monaghan J.J. 1978, MNRAS {\bf 184} 25 

\nnn Okamoto I., Katz J. and Parentani R. 1995, Class. Quantum Grav. {\bf 12} 443

\nnn Parentani R., Katz J., and Okamoto I. 1995, Class. Quantum Grav. {\bf 12} 1663

\nnn Padmanabhan T. 1990,  Physics Reports {\bf 188} No. 5, 285

\nnn Poincar\'e H. 1885, Acta Math. {\bf 7} 259

\nnn Sorkin R.D. 1981  Astrophys.J. {\bf 249} 254

\nnn Thorne, K.S. 1966, in {\it High Energy Astrophysics} eds. DeWitt C., Schatzman E. 
and V\'eron P. (Gordon and Breach, New York)

\nnn Trager S.C., Djorgovski S. and King I.R. 1993, in {\it Structure and Dynamics of
Globular Cluster} ASP Conference Series 50, eds. Djorgovski S. and Meylan G. 

\nnn Thompson J.M.T. 1979, Phil. Trans. R. Soc. Lond. {\bf 292} 1386 
\bye